\documentclass[aps,prl,twocolumn,superscriptaddress]{revtex4}
\usepackage{amsfonts}

\usepackage{graphicx,amsmath,amssymb}
\usepackage{epstopdf}
\usepackage{grffile}
\usepackage[usenames]{color}
\usepackage{indentfirst}
\usepackage{float}
\usepackage{color}
\usepackage{mathrsfs}

\begin{document}

\title{Quantum phases in a quantum Rabi
triangle}
\author{Yu-Yu Zhang}
\email{yuyuzh@cqu.edu.cn}
\affiliation{Department of Physics, Chongqing University, Chongqing 401330, China}
\affiliation{Chongqing Key Laboratory for strongly coupled Physics, Chongqing 401331, China}
\author{Zi-Xiang Hu}
\affiliation{Department of Physics, Chongqing University, Chongqing 401330, China}
\affiliation{Chongqing Key Laboratory for strongly coupled Physics, Chongqing 401331, China}
\author{Libin Fu}
\affiliation{Graduate School, China Academy of Engineering Physics, Beijing 100193, China}
\author{Hong-Gang Luo}
\affiliation{School of Physical Science and Technology, Lanzhou University, Lanzhou
730000, China}
\author{Han Pu}
\affiliation{Department of Physics and Astronomy, and Rice Center for Quantum Materials, Rice University, Houston, Texas 77251-1892, USA}
\author{Xue-Feng Zhang}
\email{zhangxf@cqu.edu.cn}
\affiliation{Department of Physics, Chongqing University, Chongqing 401330, China}
\affiliation{Chongqing Key Laboratory for strongly coupled Physics, Chongqing 401331, China}

\begin{abstract}
The interplay of interactions, symmetries and gauge fields usually leads to
intriguing quantum many-body phases. To explore the nature of emerging phases,
we study a quantum Rabi triangle system as an elementary building block for
synthesizing an artificial magnetic field. We develop an analytical approach to study the rich phase diagram and the associated quantum criticality. Of particular interest is the emergence of a chiral-coherent phase, which breaks both the $\mathbb{Z}_2$
and the chiral symmetry. In this chiral phase,
photons flow unidirectionally and the chirality can be tuned
by the artificial gauge field, exhibiting a signature of broken time-reversal symmetry.
The finite-frequency scaling analysis further confirms the associated phase transition to be
in the universality class of the Dicke model. This model can simulate a broad range of physical
phenomena of light-matter coupling systems, and may have an application in future
developments of various quantum information technologies.
\end{abstract}

\date{\today }
\maketitle

%The quantum phase transition in few body system
\textit{Introduction.--} The coupling between light and matter has brought forth a novel class of quantum many-body systems~\cite{Niemczyk,pforn,fumiki,pippan,bloch}, which is useful
in probing of a broad range of physical phenomena. The possibility
of quantum phase transition (QPT) of photons has stimulated a lot of
discussions in the Jaynes-Cummings (JC) Hubbard lattice~\cite%
{plenio,greentree,zhu} and the Rabi lattice models~\cite{flottat,zheng,schiro}.
The basic building block of such systems contains a two-level system and a
bosonic field mode, which is the simplest and the most fundamental model
describing quantum light-matter interactions~\cite{rabi,diaz19}. Usually,
the QPTs are discussed in the thermodynamical limit \cite{sachdev}. However,
the quantum Rabi model~\cite{plenio1,liu,zhang20,lv}, two-site JC lattice~%
\cite{plenio2}, and few-body systems with non-linearity~\cite{simone20} in proper limits also exhibit the similar scaling behavior of QPTs.
Such QPTs in few-body system open a window for investigating
related integrability, exotic phases and critical behaviors~\cite{plenio1,liu,zhang20,Braak,chen2012}.

The intriguing many-body phases generally arise from the interplay of strong
interactions, symmetries and external fields.
Recent years artificial gauge fields have been created for quantum platforms
with bosonic excitations, such as neutral atomic Bose-Einstein condensate (BEC) or cold quantum gases~\cite{lin2009,RMP,fu}, and photonic systems~\cite{umu,wang,Cai,roushan,bloch2012}. For example, manifestations
of artificial magnetism in quantum gases in terms of vortex nucleation have been found~\cite{RMP}, the intriguing phenomenon
of fractional quantum Hall (FQHE) physics has been predicted to occur in the JC Hubbard system
by applying an artificial magnetic field~\cite{hayward,martin2016,noh2017}. A few-body system of light-atom interactions subjected to an artificial magnetic field provides an ideal platform to investigate new quantum phases, which can be controlled conveniently by the artificial gauge fields.
%Motivated by recent remarkable progress on the implications of synthetic gauge fields for microwave photons in circuit QED ~\cite{roushan,fan,kapit,houck,umu,wang},

\begin{figure}[t!]
\centering
\includegraphics[width=\linewidth]{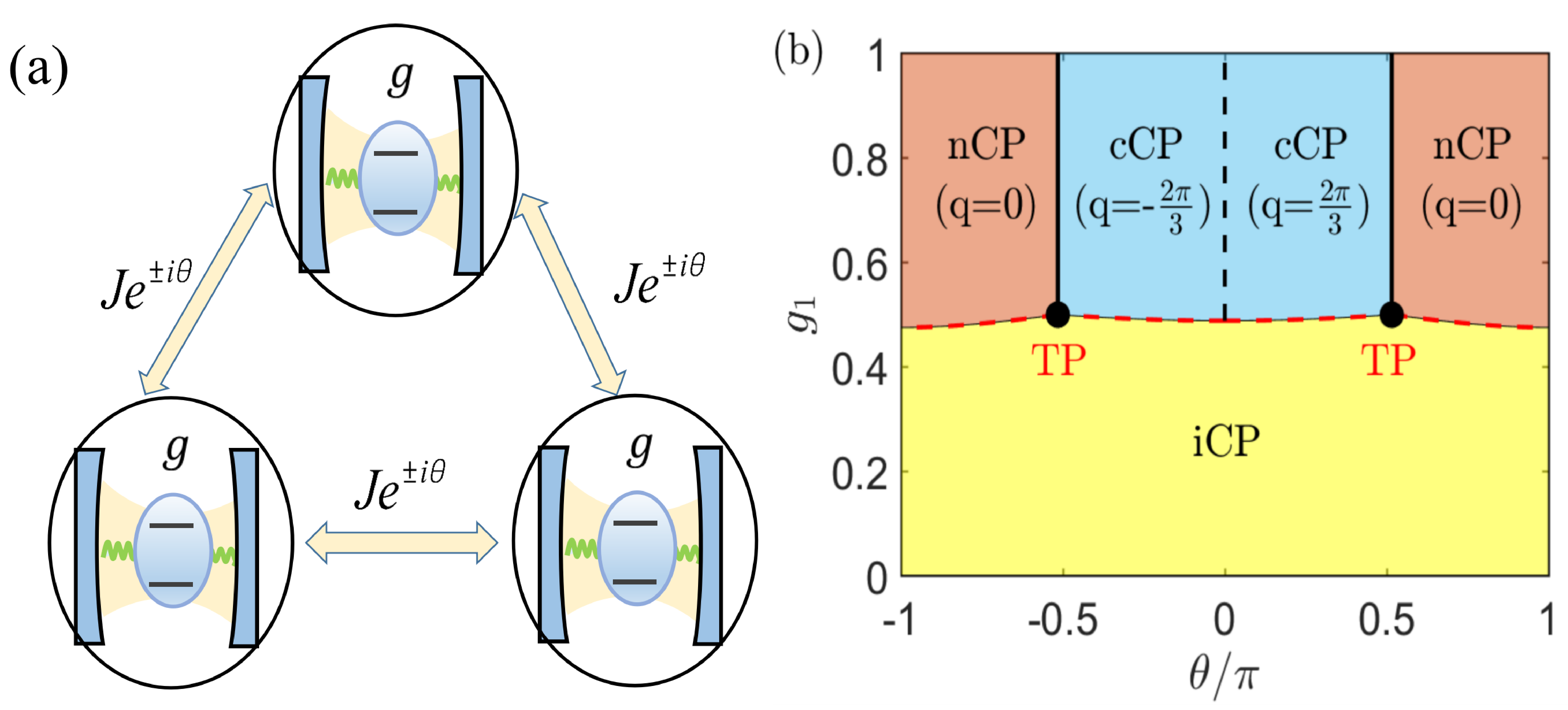}
\caption{(a)The schematic diagram of quantum Rabi triangle system with
artificial gauge field. (b) The analytic phase diagram in the $g_{1}$-$
\protect\theta$ parameter space.
The second order critical lines (red dash) from the iCP
to nCP and cCP join with the first order line (black sold) between
nCP and cCP at the triple points (TPs) (black dot). The black dash line separates the cCP
according to its chirality. In all our calculations, we set $\omega=1$ as the units for frequency, and $\Delta=50$, $J=0.05$. }
\label{fig1}
\end{figure}

%About our work
In this Letter, we study the quantum Rabi triangle (QRT), as a
fundamental unit for synthesizing a magnetic field to manipulate photons in optical cavities,
to explore the possibility of phase transitions in a few-body system.
Mean-field approximations are usually adopted in many-body systems and often yield quantitatively accurate results. This,
however, is in general no longer true in dealing with few-body systems. As such, analytic results can rarely be found in few-body systems. Remarkably, we show that exact analytic results can be found in the QRT in the infinite frequency limit (analogous to the thermodynamic limit). Using this analytic approach, we construct the phase diagram of the QRT and explore the associated phase transitions. The QRT contains three phases. An incoherent phase (iCP), analogous to the normal phase in the Dicke model, dominates the weak coupling regime. In the strong coupling regime, there exist two coherent phases: the normal coherent phase (nCP) is analogous to the superradiance phase in the Dicke model and breaks the $\mathbb{Z}_2$ symmetry; the chiral coherent phase (cCP) breaks both the $\mathbb{Z}_2$ and the chiral symmetry and is unique to the QRT without analogy in the Dicke model. The transition between nCP and cCP is of first-order and can be induced by adjusting the artifical gauge field. The transition between iCP and the two coherent phases is of second-order and, through a finite-frequency scaling, can be shown to belong to the same universality class of the superradiance phase transition in the Dicke model.

%Model--finished
\textit{Model --}  The QRT is a model of itinerant photons hopping between neighboring cavities and
interacting on-site with a two-level atom. Three cavities are placed on a
ring, see Fig.~\ref{fig1}(a), where each cavity contains a two-level atom and
is described by the quantum Rabi model. The full Hamiltonian for the QRT system reads
\begin{eqnarray}
H_{\mathrm{QRT}} &=&\sum_{n=1}^{3}H_{\mathrm{R},n}
+\sum_{n,n^{\prime }}^3J(e^{i\mathcal{\theta }}a_{n}^{\dagger }a_{n^{\prime }}+e^{-i\mathcal{\theta }%
}a_{n}a_{n^{\prime }}^{\dagger }), \label{RM}
\end{eqnarray}%
where $a_{n}^{\dagger }$ $\left( a_{n}\right) $ is the photonic creation
(annihilation) operator of the $n$-th cavity with frequency $\omega $, $Je^{\pm i\theta}$
is the hopping amplitude between cavities $n$ and $n'$,
and $H_{\mathrm{R},n}$ denotes the quantum Rabi model of the $n$-th cavity
\begin{equation}
H_{\mathrm{R},n}=\omega a_{n}^{\dagger
}a_{n}+g\left( a_{n}^{\dagger }+a_{n}\right) \sigma_{n}^{x}+\frac{\Delta }{2%
}\sigma_{n}^{z},
\end{equation}
with $\vec{\sigma}_{n}=\{\sigma_{n}^{x},\sigma_{n}^{y},\sigma_{n}^{z}\}$ the
Pauli matrix describes the two-level atom with energy gap $\Delta $ and $g$ denotes
the strength of cavity-atom coupling.
The non-zero static phase $\theta $ in the photon
hopping amplitude arises from an artificial gauge
field $A_{n,n'}$ as $\theta =\int_{r_{n}}^{r_{n^{\prime }}}A(r)dr$~\cite{hayward,martin2016}.
The gauge-invariant effective magnetic flux in the ring is $\phi =3\theta
$. This artificial gauge field can be realized by a periodic modulation of the photon hopping strength between cavities, the details of which can be found in the Supplemental Material~\cite{supp}. We will focus on the infinite-frequency limit, in which $\Delta$ is much larger than any other frequency scales in the system. This is the limit where the single-cavity  Rabi model also exhibits the superradiance phase transition~\cite{plenio1,liu,zhang20,rabi2} as in the Dicke model.

%The sysmmetry of the system
Similar to the Rabi model, the parity operator can be defined as
$\hat{P}=\prod_n \exp(i\pi\hat{N_n})$ where $\hat{N_n}=a^{\dagger}_na_n+\sigma_n^+\sigma_n^-$ is the number of excitation quanta of the $n$-th cavity. Because $[H_{\mathrm{QRT}},\hat{P}]=0$, the parity operator $\hat{P}$ is conserved and equal to $\pm1$, %Since one has $\hat{P_n^\dagger}a_n\hat{P_n}=-a_n$, and $\hat{P_n^\dagger}\sigma_n^x\hat{P_n}=-\sigma_n^x$, the QRT Hamiltonian is invariant under the transformation. The parity operator possesses two eigenvalues $\pm1$
depending on whether the total number of excitation quanta is even or odd. Besides such $\mathbb{Z}_2$ symmetry, the time-reversal symmetry (TRS) of hopping processes among three cavities is artificially broken when $\theta \neq m\pi \,(m\in \mathbb{Z})$. However, it can be recovered by implementing the chiral transformation $C_r$ which exchange the even and odd permutation ($123\leftrightarrow 321$).
Considering that the gauge field plays a critical role in the search for exotic quantum phases of matter, it can be anticipated that it will give rise to interesting properties in the QRT system. In Fig.~\ref{fig1}(b), we plot the phase diagram in the parameter space spanned by $g_1$ and $\theta$, where $g_{1}=g/\sqrt{\Delta \omega }$ is the scaled dimensionless coupling strength, and $\theta$ is restricted between $-\pi$ and $\pi$. We will now discuss the three phases in detail.

%The projection which can eliminate the high energy part
\textit{Incoherent phase --} In the weak coupling regime (i.e. small $g_1$), the number of excitation
tends to zero and no photon propagates in the cavities, we have the so-called
incoherent phase (iCP). To obtain its energy spectrum, we
first implement the Schrieffer-Wolff transformation with the unitary operator $S_n=\exp [-ig_{1}\sqrt{%
\omega /\Delta } \sigma_n^{y}\left( a_n^{\dagger }+a_n\right) ]$ on each
cavity. After neglecting higher-order terms in the limit $\Delta /\omega
\rightarrow \infty $, Hamiltonian~(\ref{RM}) becomes
\begin{eqnarray}
H_{\mathrm{iCP}} &=&\sum_{n=1}^{3}\omega a_{n}^{\dagger }a_{n}+%
\frac{\Delta }{2}\sigma_{n}^{z}+\omega g_{1}^{2}(a_{n}+a_{n}^{\dagger
})^{2}\sigma_{n}^{z}  \notag \\
&&+J(e^{i\theta }a_{n}^{\dagger }a_{n+1}+h.c.)+O(g_{1}^{4}\frac{\omega ^{2}}{\Delta ^{2}}).
\end{eqnarray}%
Because the transverse operator $\sigma_n^{x}$ is eliminated, the two atomic levels are decoupled. Thus, the low-energy effective Hamiltonian can be obtained by
projecting to the subspace of the lower atomic level $\left|\downarrow\right\rangle_n$, i.e., $H_{%
\mathrm{iCP}}^{\downarrow }=\langle\left\downarrow\right| H_{\mathrm{iCP}}|\left\downarrow\right
\rangle $.
After taking a discrete
Fourier transform $a_{n}^{\dagger }=\frac{1}{\sqrt{N}}\sum_{q}e^{inq}a_{q}^{%
\dagger }$ with the quasi-momentum $q$ taking values $0$ and $\pm 2\pi /3$, we have
\begin{equation}
H_{\mathrm{iCP}%
}^{\downarrow}=E_0+\sum_{q}\omega _{q}a_{q}^{\dagger }a_{q}-\omega
g_{1}^{2}(a_{q}a_{-q}+a_{q}^{\dagger }a_{-q}^{\dagger })\,,\label{Hicp}
\end{equation}
where  $E_{0}=-3\Delta/2-3\omega
g_1^2+3(\omega+J)g_1^2\omega/\Delta$ is a constant, and $\omega
_{q}=\omega -2\omega g_{1}^{2}+2J\cos (\theta -q)$ (see Supplemental
Material~\cite{supp}). Hamiltonian (\ref{Hicp}) is quadratic in photon operators and hence can be diagonalized using the Bogoliubov transformation~\cite{supp}. The diagonalized Hamiltonian takes the form
$H_{\mathrm{iCP}}^{\downarrow}=\sum_{q}\varepsilon _{q}a_{q}^{\dagger }a_{q}+E_{g}$,
where $E_{g}=\sum_{q}(\varepsilon _{q}-\omega _{q})/2+E_{0}$ is the ground-state energy,
and the photon dispersion is given by
\begin{equation}  \label{excited}
\varepsilon _{q}=\frac{1}{2}[\sqrt{(\omega _{q}+\omega _{-q})^{2}-16\omega
^{2}g_{1}{}^{4}}+\omega _{q}-\omega _{-q}].
\end{equation}
The excitation spectra $\varepsilon _{q}$ with the momentum $q=0,\pm2\pi/3$ decreases to zero
as the coupling strength increases to a critical value (see Fig.$1$ in the supplementary material \cite{supp}).

\begin{figure}[tbp]
	\includegraphics[trim=70 100 190 20,width=0.9\linewidth]{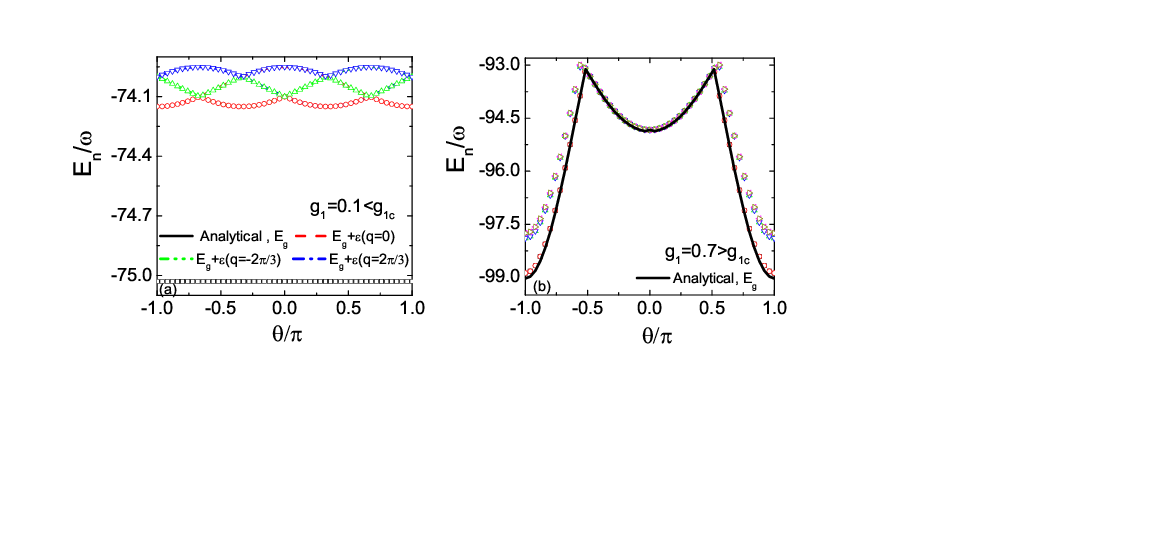}
	\caption{(a) Energy spectrum in the iCP phase with $g_1=0.1$. Curves represent analytic result. Symbols correspond to the lowest 4 eigenenergies numerically obtained from ED.
(b)  Energy spectrum in the nCP and cCP phases with $g_1=0.7$. Black curve corresponds to analytic ground-state energy. Symbols correspond to the lowest 6 eigenenergies numerically obtained from ED.
		 Other parameters are the same as in Fig.~\ref{fig1}. The two peaks are located at $\pm \theta_c=\pm 0.516\pi$.}
	\label{energy}
\end{figure}

Figure~\ref{energy}(a) shows the analytical ground-state energy and the first few excited-state energies for the iCP phase,
which agree well with numerical results obtained from exact diagonalization (ED) of the original Hamiltonian~(\ref{RM}).
It is observed that the iCP is a gapped phase with non-degenerate ground state and there exist energy-level crossings in excited states.
It should be noted that the ground state has even parity. This can be understood from the fact that at $g_1=0$, there are no photons and all the atoms are in the lower level $|\downarrow \rangle$ in the ground state, which clearly has an even excitation number $0$.

\textit{Coherent phases --} In the strong coupling regime, there exist two coherent phases, in which the cavity field is macroscopically populated~\cite{plenio1}. To obtain the effective Hamiltonian, we first shift the cavity operator as  $a_{n} \rightarrow a_{n} +\alpha
_{n} $ with the complex displacement $\alpha _{n}$. With the displaced operator, the QRT Hamiltonian takes the form
\begin{eqnarray}\label{hamcp}
H_{\mathrm{CP}}&=&\sum_{n}\omega a_{n}^{\dagger }a_{n}+\frac{\Delta
	_{n}^{\prime }}{2}\tau ^{z}_{n}+g^{\prime}_n\left( a_{n}^{\dagger
}+a_{n}\right) \tau ^{x}_{n}  \notag \\
&&+Ja_{n}^{\dagger }(e^{i\theta }a_{n+1}+e^{-i\theta }a_{n-1})+V_{off}+E_{0},
\end{eqnarray}%
where $\Delta _{n}^{\prime }=\sqrt{%
	\Delta ^{2}+16g^{2}A_{n}^{2}}$ is the renormalized energy gap, and
$g^{\prime}_n=g\Delta/\Delta _{n}^{\prime }$ the effective coupling strength. Here, $\tau_{n}^{z}=\Delta /\Delta _{n}^{\prime }\sigma_{n}^{z}+4gA_{n}/\Delta _{n}^{\prime }\sigma_{n}^{x}$ is the transformed Pauli matrix. The off-diagonal term $V_{off}$
and the energy constant $E_{0}$ are given in the the Supplemental Material~%
\cite{supp}. A proper choice of the displacement $\alpha_n$ leads to the  vanish of $V_{off}$ and, as a result, Hamiltonian~(\ref{hamcp}) has the same structure as Hamiltonian~(\ref{RM}) with the rescaled frequency $\Delta _{n}^{\prime }$ and coupling strength $g^{\prime}_n$. Therefore, by employing the same procedure used to derive $H_{\mathrm{iCP}}$, we obtain
the effective Hamiltonian in the coherent phases by projecting to the spin subspace $|\downarrow \rangle $
\begin{eqnarray}
H_{\text{CP}}^{\downarrow }&&=\sum_{n=1}^{3}\omega a_{n}^{\dagger }a_{n}-\frac{g^{\prime 2}_n}{\Delta _{n}^{\prime }}
\left( a_{n}^{\dagger }+a_{n}\right) ^{2}-\frac{\Delta _{n}^{\prime }}{2} \nonumber\\
&&+Ja_{n}^{\dagger }(e^{i\theta}a_{n+1}+e^{-i\theta }a_{n-1})+E_{0}.
\label{hmccp2}
\end{eqnarray}

\begin{figure}[tbp]
	\includegraphics[trim=50 100 180 20,width=0.8\linewidth]{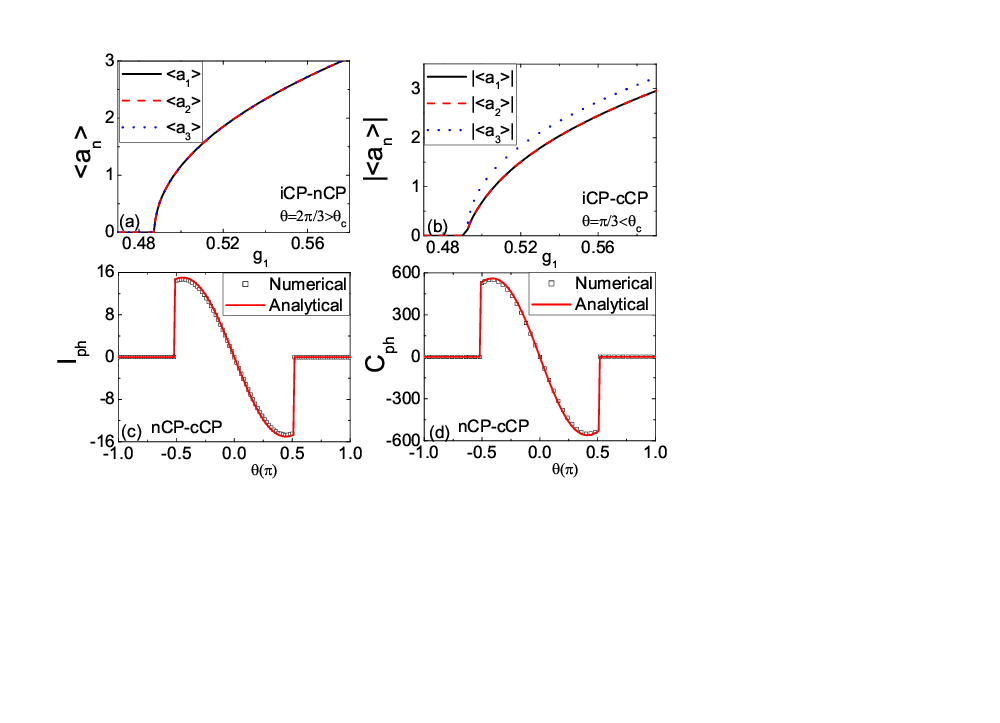}
	\caption{(a) The order parameter $\langle a_n\rangle$ as a function of the scaled coupling
		strength $g_{1}$ for the iCP-nCP transition with $\theta=2\pi/3>\theta_c$. (b) $|\langle a_n\rangle|$ as a function of $g_1$ for the iCP-cCP transition with $\theta=\pi/3<\theta_c$.
(c) Photon current $I_{\mathrm{ph}}$ and (d) the expected value of the chirality operator $C_{\mathrm{ph}}$
in the ground state as a function of the hopping phase $\theta$ for the nCP-cCP transition with $g_1=0.7>g_{1c}$.
Other parameters are the same as in Fig.~\ref{fig1}.}
	\label{alpha}
\end{figure}

Diagonalizing the above quadratic Hamiltonian, we obtain two coherent phases (see Fig.~\ref%
{fig1}(b)):\newline
(i) \textbf{normal-coherent phase (nCP)}. The nCP occurs for $|\theta|>\theta_c$, where $\theta_c$ is a critical value for the phase of the photon hopping amplitude (see below). In the nCP, the ground state features $q=0$ which indicates that photons have zero quasi-momentum, and $\alpha_n$ can be taken to be real with the explicit expression~\cite{supp}
$\alpha_{n}=\sqrt{\frac{g^{2}}{(\omega +2J\cos \theta )^{2}}-\left(\frac{\Delta}{4g}\right)^2}$ independent of $n$.
The photon dispersion is given by
\begin{equation}
\varepsilon _{q}=\frac{1}{2}(\omega _{q}^{\prime}-\omega _{-q}^{\prime})+%
\sqrt{(\omega _{q}^{\prime}+\omega
	_{-q}^{\prime})^{2}-16g^{\prime 4}_n/\Delta _{n}^{\prime 2}}.
\end{equation}
where $\omega_{q}^{\prime}=\omega-2g_n^{\prime
	2}/\Delta_n^{\prime}+2J\cos(\theta-q)$. Furthermore, the ground state is two-fold degenerate as a result of the $\mathbb{Z}_2$ symmetry breaking. This two-fold degeneracy can be seen from the ED numerical results presented in Fig.~\ref{energy}(b), where the left and right parts of the curve represent the nCP. In Fig.~\ref{alpha}(a) we show how the order parameters $\langle a_n \rangle$ varies as a function of the coupling strength $g_1$. In increasing $g_1$, the system enters from the iCP (where $\langle a_n \rangle=0$) to the nCP, and the order parameter grows from zero, indicating a second-order phase transition. In nCP, $\langle a_n \rangle$ are the same for all three cavities. Figure~\ref{alpha}(a) only shows one of the two degenerate ground-state solutions for nCP. The order parameter takes a minus sign in the other solution. \newline
(ii) \textbf{chiral-coherent phase (cCP)}. The cCP, which occurs when $|\theta|<\theta_c$, features finite photon quasi-momentum $q=\pm2\pi/3$. Here the displacement $\alpha_n$ is in general complex and $n$-dependent. The middle part of Fig.~\ref{energy}(b) between the two cusps, denoting the position of $\pm \theta_c$, represent the ground-state energy of cCP. The ED results also clearly show that the ground state has 6-fold degeneracy. This is because, in addition to the $\mathbb{Z}_2$ symmetry, the cCP also breaks the chiral symmetry resulting in a unidirectional photon current. Fig.~\ref{alpha}(b) shows how the magnitude of the order parameter vs. $g_1$ when the system enters from iCP to cCP. One can again see a second-order phase transition. However, different from nCP, the order parameters in cCP are $n$-dependent and are in general complex. For the example shown in the figure, the phase angles for $\langle a_{1,2,3} \rangle$ are $\pi$, $-0.11126$ and $0.11126$, respectively, and are nearly insensitive to the value of $g_1$. Here we only show one of the six degenerate ground-state solutions. In the other solutions, the order parameters take cyclic permutations and/or take a minus sign. To better characterize the photon current and the chirality, we define the photon current operator as
$I_{\mathrm{ph}}=i\left[(a_{1}^{\dagger }a_{2}+a_{2}^{\dagger
}a_{3}+a_{3}^{\dagger }a_{1})-h.c.\right],$
%In the momentum space, one obtains $I_{\mathrm{ph}}=\sum_{q}a^{%
%\dagger}_qa_q\sin(q)$, which predicts the smooth function of the photon current.
% However, as the two-body operator, it can not reflect the chirality of the system.
analogous to the continuity equation in classical systems.
Moreover, in analogy with the spin chiral operator via Pauli matrix $C=\frac12\sum_{<ijk>}\vec{\sigma}_{i}\cdot(\vec{\sigma}
_{j}\times \vec{\sigma}_{k})$ ~\cite{wen1989}, the photon chiral operator can be defined as
$C_{\mathrm{ph}}=-2i\sum_{<ijk>}\varepsilon _{ijk}a_{i}a_{j}^{\dagger}(n_{k}-1/2)$ ($\varepsilon _{ijk}$ is Levi-Civita tensor) with help of linearized spin-wave transformation $\sigma_{i}^{-}=a_{i}$, $\sigma _{i}^{+}=a_{i}^{\dagger }$ and $\sigma_{i}^{z}=2a_{i}^{\dagger }a_{i}-1$~\cite{zheng,mattis1988}. Similar to the spin system, the photon chiral operator is odd under either the chiral transformation $C_r^{-1}C_{\mathrm{ph}}C_r=-C_{\mathrm{ph}}$, or the TRS transformation. Meanwhile, the photon current operator has the same properties of the symmetries. In Fig.~\ref{alpha}(c) and (d), we show $I_{\rm ph}$ and $C_{\rm ph}$, respectively, as functions of $\theta$. One can see that these two quantities are zero for nCP and finite for cCP, except at $\theta=0$ where TRS is recovered in the Hamiltonian.

%The critical line obtained by checking gap close
\textit{Quantum criticality an phase boundaries --} As mentioned above, the transition from the iCP to either coherent phases is of second-order and is induced by varying the coupling strength $g_1$. The critical coupling strength $g_{1c}$ can be obtained from the excitation spectra $\varepsilon _{q}$ in Eq.~(\ref{excited}) --- $\varepsilon _{q}$ must vanish at $g_{1c}$,
yielding
\begin{equation}
g_{1c}(q)=\sqrt{\frac{1+\frac{4J}{\omega }\cos \theta \cos q+\frac{4J^{2}}{%
\omega ^{2}}\cos (\theta +q)\cos (\theta -q)}{4(1+\frac{2J}{\omega }\cos
\theta \cos q)}}.
\end{equation}%
The transition between the two coherent phases, by contrast, is of first-order, features discontinuous jump in the order parameter, and is induced by varying the effective magnetic flux $\theta$. Using the analytic expressions of the ground-state energy for nCP and cCP, we obtain the critical value $\theta_c$ as~\cite{supp}
\begin{eqnarray}
\theta _{c}=\cos^{-1}\left(-\frac{2 J}{\sqrt{8 J^2+\omega ^2}%
	+\omega }\right),
\end{eqnarray}
and the phase boundary between nCP and cCP occurs at $\pm \theta_c$. Note that $\theta_c$ is independent of $g_1$.

These results allow us to construct the phase diagram presented in Fig.~\ref{fig1}(b). There two triple points (TPs) in the phase diagram, at which all three phases co-exist. The TPs are located at $(g_{tc},\pm \theta_c)$ where the value of $g_{tc}$ can be obtain from $g_{1c}(q=0)=g_{1c}(q=\pm 2\pi/3)$, which yields
\begin{eqnarray}
g_{tc}&=&\frac{1}{2} \sqrt{\frac{3}{2}-\frac{\sqrt{8 J^2+\omega ^2}}{2
\omega }}.
\end{eqnarray}
%In the limit $J/\omega\rightarrow 0$, the TPs are located at $%
%\theta_c=\pm\pi/2$ and $g_{tc}=1/2$.%, which reduces to the critical coupling strength of the single-cavity quantum Rabi model~\cite{plenio1}.

%How to solve the equation to obtain the coherent phase

%shows the behavior of the ground-state energy $E_g$ as a function of the phase $\theta$, showing the two representative cases for the NCP and CCP respectively. As the numerical convergence of the ground-state energy depends on the bosonic Hilbert space, one needs to use the truncated photon number $N_{tr}$ large enough to obtain sufficient numerical accuracy. Fortunately, $E_g$ obtained analytically is consistent with the numerical ones with the truncated photon number up to $N_{tr}=35$, which verifies the validity of our solutions. The ground-state energy is discontinuous at $\theta_c=\pm0.498$, locating the first-order transition from the NCP to CCP.

%there is no photon current $I_{\mathrm{ph},0}$ in the ground state, and thechirality $C_{\mathrm{ph},0}$ is also zero in Fig.~\ref{current phase} (a-b).
%For the first-excited state, photons can be triggered to propagate in
%three cavities by the gauge field. Both of $I_{\mathrm{ph},1}$ and $C_{\mathrm{ph},1}$
%exhibit discontinuous jumps
%at the hooping phase $\theta=0, \pm2\pi/3$, where the
%energy level-crossing occurs in Fig.~\ref{energy}(a).
%It ascribes to the excitation energy $\varepsilon_q$
%with different momentum dependent on $\theta$.

\begin{figure}[tbp]
\includegraphics[trim=15 100 30 20,scale=0.8]{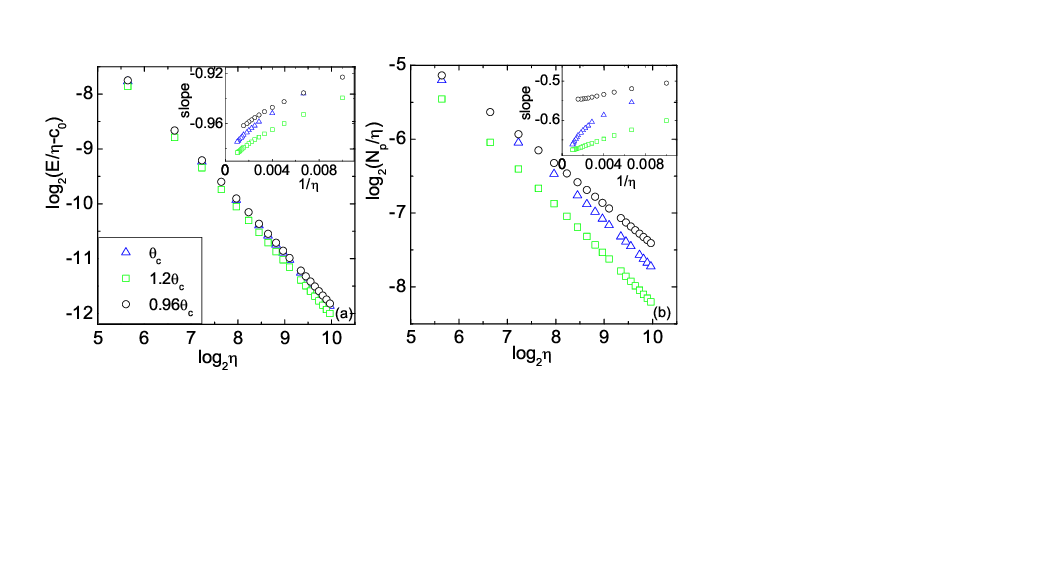}
\caption{Scaling of the ground-state energy (a) and photon number (b) as a
function of $\protect\eta$ at the critical point for different gauge field
phase $\protect\theta=\protect\theta_c$, $1.2\protect\theta_c$ and $0.96%
\protect\theta_c$ for continuous QPTs of the iCP-nCp and iCP-cCP transitions. The insets show the corresponding slope versus $1/\protect%
\eta$.}
\label{scaling1}
\end{figure}

\textit{Universal scaling --} The QRT Hamiltonian can exhibit a scaling relation for finite values of $\Delta/\omega$ as a consequence continuous QPTs in the thermodynamic limit. The universal scaling of the QPTs can be
characterized by the critical exponents for finite values of $%
\eta=\Delta/\omega$. Figure~\ref{scaling1} illustrates the finite-$\eta$
scaling of the ground-state energy and the average photon number obtained by
numerical diagonalization in the critical regime.
% The accuracy of the exponents are limited by the available numerical date.
In the limit $\eta\rightarrow\infty$, the scaled ground-state energy $%
E_g/\eta$ obtained analytically at the critical point approaches $%
c_0=-3\omega/2$. To show the leading finite-$\eta$ corrections, we calculate
$E_g/\eta-c_0$ versus $\eta$ at the critical value $g_{\texttt{1c}}$ on a log-log scale in
Fig.~\ref{scaling1} (a) when the system
undergoes the iCP-cCP QPT with $\theta=1.2\theta_c$ and the iCP-nCP QPT with $\theta=0.96\theta_c$, respectively. The corresponding slope of the curves in the large-$%
\eta$ regime gives a universal exponent $-1$ for both QPTs. Meanwhile, a power-law
behavior of the photon number $N_p=\sum_n\langle a_n^{\dagger}a_n\rangle$
exists at large $\eta$ as shown in Fig.~\ref{scaling1} (b). The corresponding finite-$%
\eta$ exponent extracted from the curves converges to $-0.667$ as shown in the
inset. To conclude, we find that the scaling exponents for the ground-state energy and
the average photons number are universal, giving two power law expressions
as $E_g/\eta-c_0\propto \eta^{-1}$ and $N_p/\eta\propto \eta^{-2/3}$ for both the iCP-nCP and the iCP-cCP transitions, belonging to the same universality class of the Dicke model~\cite%
{lambert,chen2} and the single-site Rabi model in the infinite-frequency limit~\cite%
{plenio1,liu}.

\textit{Conclusion --} We present an exact analytic solution to the quantum Rabi
triangle system as a basic building block for exploring strongly correlated
physical phenomena. We identify the quantum phases and the transitions among them. In particular, there is an exotic chiral coherent phase that has no analog in the single-cavity Dicke or Rabi models. The cCP breaks both the $\mathbb{Z}_2$ and the chiral symmetry, featuring a persistent unidirectional photon current in its ground state. The current and the chirality can be tuned by the phase of the inter-cavity photon hopping amplitude, which plays the role of an artificial magnetic flux.

Our study advances the field of strongly correlated photons in light-atom coupled system.
Studying the quantum phases in this few-body system under the introduction
of an artificial magnetic field
would open intriguing avenues for exploring their connection to strongly correlated photons in
two-dimension lattices system~\cite{flottat,zheng,schiro}.
Moreover, an implementation of the system considered in this Letter is an exciting prospect
for the future and may be applicable in future developments of various quantum information technologies.
One has proposed an application of the Mott state in the JC Hubbard lattice for implementing quantum information processing~\cite{noh2017}.
One could hope to implement cluster state quantum computing related to extension of the quantum Rabi triangle system coupled many resonators for strong atom-resonator coupling. Our studies also shed new light in quantum simulation of artificial magnetic field in ultracold bosonic atoms~\cite{RMP}.

The authors thank Qing-Hu Chen for useful discussions. This work was
supported by NSFC under Grants No. 12075040, No. 11804034, No. 11874094, No. 12047564, No. 11834005,
No. 11974064, No. 11725417, No. 12088101, No. 11834005 and No. 12047501, NSAF under Grant No. U1930403, and Chongqing NSF under Grants No. cstc2020jcyj-msxmX0890 and No. cstc2018jcyjAX0399, Fundamental Research Funds for the Central Universities Grant No. 2021CDJQY-007. H.P. acknowledges support from the US NSF and the Welch Foundatioin (Grant No. C-1669).

\end{document}